# Scanning-gate microscopy of semiconductor nanostructures: an overview

F. MARTINS[a], B. HACKENS[a], H. SELLIER[b], P. LIU[b], M.G. PALA[c], S. BALTAZAR[c], L. DESPLANQUE[d], X. WALLART[d], V. BAYOT[a,b] and S. HUANT[b,e]

[a] *IMCN, pôle NAPS, Université catholique de Louvain, B-1348 Louvain-la-Neuve, Belgium*
[b] *Institut Néel, CNRS & Université Joseph Fourier, BP 166, F-38042 Grenoble, France*
[c] *IMEP-LAHC, UMR 5130, CNRS/INPG/UJF/UdS, Minatec, Grenoble, France*
[d] *IEMN, Cité scientifique, Villeneuve d'Ascq, France*
[e] *Email address : serge.huant@grenoble.cnrs.fr*

This paper presents an overview of scanning-gate microscopy applied to the imaging of electron transport through buried semiconductor nanostructures. After a brief description of the technique and of its possible artifacts, we give a summary of some of its most instructive achievements found in the literature and we present an updated review of our own research. It focuses on the imaging of GaInAs-based quantum rings both in the low magnetic field Aharonov-Bohm regime and in the high-field quantum Hall regime. In all of the given examples, we emphasize how a local-probe approach is able to shed new, or complementary, light on transport phenomena which are usually studied by means of macroscopic conductance measurements.

## 1. Introduction

Scanning-gate microscopy (SGM) consists in scanning the electrically polarized tip of a cooled atomic-force microscope (AFM) above a semiconductor device while simultaneously mapping the conductance changes due to the tip perturbation. Since its introduction in the late nineties [1-3], SGM has proven powerful to unravel the local details of electron behavior inside modulation-doped nanostructures whose active electron systems are, in contrast to surface electron-systems [4], not accessible to scanning-tunneling microscopy (STM) because they are located too deep below the surface. After a brief description of the technique as well as of some instructive achievements found in the literature [2,3,5-9], this paper contains a selection of our own research dealing with the imaging of electron transport across GaInAs quantum rings (QRs) both in the (low-field) Aharonov-Bohm (AB) [10-13] regime and in the (high-field) quantum Hall [14] regime.

## 2. The technique of scanning-gate microscopy

The principle of scanning-gate microscopy is sketched in Fig. 1. A semiconductor nanostructure (here ring-shaped) hosts a two-dimensional electron gas (2DEG) buried at typically a few tens of nanometers below the free surface. A current is injected into the device in order to measure its conductance. A polarized ($V_{tip}$) AFM tip is scanned at some 20 nm over the device to perturb as locally as possible its conductance whose induced changes are mapped as function of the tip position. The whole setup is plunged into a cryostat to operate at low temperature. Optionally, a perpendicular magnetic field can be applied. This requires the use of magnetic-free cryogenic step-motors [15] and piezo-scanners to displace the nanostructure under the tip at low temperature. The microscope in Grenoble operates at 4 K and in a magnetic field up to 11 T, whereas that in Louvain-la-Neuve, which has been developed very recently, can operate in a dilution fridge down to temperatures below 100 mK and in a field as high as 17 T.

SGM experiments have several difficult steps to deal with. The most critical one perhaps is to locate without damage (either electrical or mechanical) the region of interest under the tip in a top-loading microscope which has no optical access to the user. This is achieved by patterning a set of arrows and other "traffic signs" around the device. Locating these signs by recording the sample profile in the AFM tapping mode serves as a guide to drive the tip towards the device.

Measuring the sample topography requires measuring the AFM cantilever deflection by some means. Instead of using an optical method as commonly done in AFM, it is wiser to use here a light-free setup such as e.g. a quartz tuning-fork originally introduced for near-field scanning optical microscopy [16-18] and extended later to other scanning-probe microscopy (see e.g. [19,20]). This is because semiconductor materials are photosensitive and illuminating them can induce undesired persistent changes in their properties such as the carrier density in the active 2DEG [10,11]. For the purpose of SGM, either a metallic tip or a commercial conductive AFM cantilever can be electrically anchored on one metallic pad of the tuning-fork. A further advantage of using a tuning fork is that both the topographic profile and conductance images can be recorded simultaneously. A drawback however is that the tuning fork is stiff, much stiffer than a standard AFM cantilever, and special care must be taken not to apply a too large tapping force [16-18] on the sample which could irreversibly damage it. Alternatively, soft piezoelectric cantilevers can be used [1-3].

Once the device has been located under the tip, the latter is lifted at some tens of nanometers above the surface and scanned in a plane parallel to the 2DEG (the tip does not follow the topography). Therefore, all subsequent SGM measurements are carried out without contact to the surface. However, it may happen that drifts occur during SGM imaging, especially if parameters like temperature or magnetic field are varied. In this case, additional topographic images are acquired to compare with the initial topography and possibly adjust the sample position.

Apart from the topographical behavior of the tip, its "electrical behavior" is also crucial in SGM. This is discussed in the next section.

## 3. Some tip artifacts in SGM

Experimentally evaluating [21] or theoretically computing the tip potential seen by the electrons probed in SGM is a complex matter. The electrostatic AB effect introduced later in this paper (section 5) indicates that the tip potential is strongly screened by the electron system. This can be simulated by solving in a self-consistent way the Schrödinger and Poisson equations governing the properties of the electron system coupled to the tip potential, which remains to be done to the best of our knowledge.

Here, our goal is less ambitious and we simply wish to put forward some artifacts which we encountered during our own studies and which can be circumvented with some caution.

The first one is related to the magnitude of the bias applied to the tip. Obviously, applying too high a voltage will strongly couple the tip to the electrons and modify their properties. This can be simulated without resorting to self-consistent simulations as done in [11] where we found that for small tip potentials the conductance image essentially reflects the electronic local density-of-states (LDOS), as mentioned below in section 5. However, applying too high a tip potential progressively introduces spurious features in the conductance image that are not present in the LDOS (see Fig. 4 in [11]). Experimentally, the low tip-potential regime is maintained as long as the magnitude of the imaged tip-induced features, such as the "inner fringes" discussed later in section 5 (Fig. 3), increases linearly with the tip bias [11]. Beyond this regime, this magnitude tends to increase sub-linearly, or to saturate, and spurious features, not seen at low bias, appear. This regime must preferentially be avoided.

Another artifact is related to possible distortions of the tip, which may happen after thorough scanning of the sample or inadvertent tip contact to the surface. It results in distortions of the SGM images, which can also be simulated (see Fig. 2 in [12]). For large distortions, it may even happen that the respective roles of the electron system and of the tip are reversed, as suggested by the SGM image in Fig. 2b. Here the tip has been damaged during scanning of the sample in such a way that the tip potential is imaged rather than the intrinsic properties of the electron system. Fortunately, a strong indication that the tip has been damaged and that the SGM data are consequently not reliable is given by the strongly distorted topographic image in Fig. 2a, where the ring geometry can hardly be recognized in contrast with topographic images acquired with good tips (see e.g. Fig. 1a in [10]).

## 4. A summary of some achievements found in the literature

Here, we give a brief summary of some achievements of SGM, which have given strong impulse to this imaging technique. We refer the reader to the cited literature for more information.

One spectacular achievement is the imaging of coherent electron flow from quantum point contacts (QPCs) patterned in high-mobility GaAs/GaAlAs 2DEGs. In QPCs, the conductance is quantized in units of $\frac{2e^2}{h}$ ($e$: electron charge, $h$: Planck's constant), each plateau corresponding to a conductance channel propagating through the QPC with near unity transmittance [22]. SGM [2] has been able to image the three lowest conductance modes in real space and has shown that each mode contributes to a number of spatial electron "strands" that is indexed in correspondence with its own index, i.e., the lowest mode contributes one strand; the second mode contributes two strands, and so on. Each strand has a well-defined modal structure that agrees well with the corresponding squared wavefunction [2]. In addition, the electron flow images are decorated with fine undulations at half the Fermi wavelength that are oriented perpendicularly to the flow and extend to micrometers away from the QPC [2,3]. They are due to coherent backscattering of electron waves between the tip depleted region and the QPC at short distance from the QPC, or between the tip and localized scattering centers at larger distances. Their occurrence confirms that the electron flow is imaged in the coherent regime of transport (the phase coherence length is of the order of a few μm in these systems at low temperature). Another interesting feature of the coherent flow from a QPC is that each electron strand splits in an increasing number of smaller ramifications at increasing distances from the QPC. These are explained as being due to focusing of the electron paths by undulations in the background potential [3].

More recent work has revealed that the larger the 2DEG mobility, the farther from the QPC does the above branching occur [5], in agreement with an increased mean free path. In the highest mobility samples, the fine decoration at half the Fermi wavelength mentioned above does not survive at large distances from the QPC as a consequence of a cleaner background potential landscape [5].

The imaging of coherent flow from QPCs summarized here has given strong confirmation that the SGM technique is a powerful tool to study, image and possibly control electron transport in mesoscopic systems. Some other spectacular achievements include the imaging of single-electron states in the Coulomb blockade regime of Carbon-nanotube quantum dots [6], lithographically patterned dots [7], self-organized dots [8] or multiple-dots along nanowires [9], edge states and localized states in the quantum Hall regime [23-26], scarred wavefunctions in open quantum billiards [27], and QRs [10-14], to list a few. Note that if SGM has been primarily applied to modulation-doped semiconductor structures, it has also been very successful with other systems such as for example Carbon nanotubes [6].

### 5. Imaging of quantum rings in the low-field Aharonov-Bohm regime

An open QR in the coherent regime of transport sees its conductance peaking when electron waves interfere constructively at the output contact and decreasing to a minimum for destructive interferences. Varying either the magnetic flux captured by the QR or the electrostatic potential in one arm, e.g., by approaching the SGM biased tip, allows the interference to be tuned. This gives rise to the well-known magnetic [28] and electrostatic [29] AB oscillations in the ring conductance. Note that strictly speaking, only scalar or vector potentials - associated with the electric or magnetic fields – are applied to the electron waves. In experiments on mesoscopic quantum rings, however, there is a common use of the term "Aharonov-Bohm effect" even though electric and magnetic fields are applied to electron waves.

We have shown some time ago that these archetypical interference phenomena can be imaged in real space by the SGM technique [10]. The electrostatic AB effect gives rise at low magnetic field to a well-developed fringe pattern in the (filtered) conductance image of GaInAs-based QRs in the coherent regime of transport (the mean-free path and coherence length in GaInAs at 4.2 K are 2 µm and 1 µm, respectively) when the tip scans outside the QR. This outer pattern is mainly concentric with the ring geometry, as can be seen in the sequence of images shown in Fig. 3a-c obtained at different voltages applied to the tip. This pattern is more clearly seen on the left part of the ring, which is supposed to be due to a ring asymmetry. The qualitative interpretation (see below for a more quantitative approach) in terms of a scanning-gate-induced electrostatic AB effect is that as the tip approaches the QR, either from the left or right, the electrical potential mainly increases on the corresponding side of the QR. This induces a phase difference between electron wavefunctions travelling through the two arms of the ring, and/or bends the electron trajectories, which produces the observed pattern. Modifying the magnetic field strength contributes another phase term through the magnetic AB effect and displaces the whole pattern with respect to the QR. This displacement is periodic in magnetic field strength with the same periodicity as the AB oscillations seen in the magneto-conductance [10], which gives further support to our interpretation in terms of AB effects.

Here, to reinforce the interpretation in terms of electrostatic AB effect, we wish to comment in more detail the tip bias effect on the conductance sketched in the images of Fig. 3a-c. A closer inspection of these images shows that the tip bias affects the position of the outer fringes. When $V_{tip}$ is raised, the outer concentric fringes are shifted away from the ring region. This translation effect is clearly evidenced in Fig. 3d,e, which depict a sequence of averaged profiles versus $V_{tip}$ for two regions labelled α and β (Fig. 3c) located on both sides of the QR. These regions were chosen for their fringe visibility. We interpret this observation as the result of the electrostatic AB effect discussed above. This effect changes the interference condition between the electrons flowing in the two arms of the ring, and thus the transmission through the device. The displacement of the outer fringes combined with the insensitivity of their amplitude to the tip bias, are clear indication that they originate from a pure interference phenomenon. Therefore, in this regime, we can directly compare the phase shifts induced by the tip electric field and by the external magnetic field. In our experiment a phase shift of π is obtained for a tip bias variation ($\Delta_{tip}$) of 1.75 V (Fig. 3d), whereas in a magnetic field the same shift is obtained with a 13 mT increment.

In addition we can estimate the lever arm of the gating effect, *i.e.* the *ratio* of the electron gas potential to the applied tip bias. In a simple model where a potential difference $\Delta V_{gas}$ is applied between the two arms of length $L$, the phase shift difference $\Delta\Phi$ of the electron wavefunctions writes: $\Delta\Phi = \pi \dfrac{e\Delta V_{gas}}{E_F} \dfrac{L}{\lambda_F}$. Using $E_F$ = 100 meV, $\lambda_F$ = 20 nm, and $L$ = 600 nm, which are realistic values for the QR imaged in Fig. 3a-c, we obtain $\Delta V_{gas}$ = 3.3mV for a phase shift of $\pi$, and therefore a lever arm of approximately 0.002. This small value is a consequence of the rather large distance of around 50 nm between the tip and the active layer and likely reflects a large screening by the 2DEG.

Turning back to Fig. 3a-c, we note that the conductance images also exhibit a complex pattern when the tip scans directly over the QR region. These inner fringes have been discussed in our previous papers and linked to the electron-probability density in the QR [11-13]. A detailed analysis based on quantum mechanical simulations of the electron probability density, including the perturbing tip potential, the magnetic field, and the presence of randomly distributed impurities, is able to reproduce the main experimental features and demonstrates the relationship between conductance maps and electron probability density maps. An example of such a relationship is shown in Fig. 4 in the case of a realistic QR perturbed by positively charged impurities. Although the impurities distort the LDOS, this distortion is reflected back in the conductance image in such a way that the conductance map can still be seen as a mirror of the electronic LDOS. As seen in Fig. 4, both the LDOS and conductance images tend to develop radial fringes, which are mostly, but not entirely, anchored to the impurity locations.

The discussion above shows that SGM can be viewed as the analogue of STM [4] for imaging the electronic LDOS in open mesoscopic systems buried under an insulating layer or the counterpart of the near-field scanning optical microscope that can image the photonic LDOS in confined nanostructures, provided that the excitation light source can be considered as point-like [30] such as in active tips based on fluorescent nano-objects [31,32].

## 6. Imaging of quantum rings in the quantum Hall regime

The above AB related phenomena occur at low magnetic field and they all disappear at higher field. This is because at high field, the cyclotron radius shrinks below the width of the QR arms and openings, and electrons tend to propagate along the edges of the device. As a consequence, the magnetoresistance of the QR exhibits plateaus at high magnetic field, as shown in Fig. 5a. These plateaus, which appear at quantized values of the resistance, are signature of the quantum Hall effect (QHE) [33].

This QHE regime is also very interesting to study by means of SGM. An example of such a (preliminary) study is given in Fig. 5b-d. This particular QR is supplied with lateral gates (insert in Fig. 5a), in contrast with the previous one, so that it is possible to monitor *in situ* the carrier concentration, i.e. the Landau level filling factor in the 2DEG at some field strength. In the (unfiltered) conductance SGM images shown in Fig. 5b,c, which are obtained at very low temperature (100 mK), in a field of 8 T and at gate voltages $V_g$ of 6.45 V and 6.85 V, respectively, we observe very narrow concentric oscillations of resistance around the QR that are discussed briefly below. Fig. 5d shows an SGM image also obtained at B= 8 T, but at a larger gate voltage $V_g$ = 7.20 V. In these conditions, the resistance of the QR is in the quantized regime. On this SGM image, we observe that the contrast completely vanishes, as has been reported by other groups, see e.g. [25]. This is due to the fact that the current through the device is carried entirely by incompressible edge states [33], and is therefore insensitive to local variations of potential caused by the tip.

As can be seen in Fig. 5a, the QR magnetoresistance displays strong reproducible fluctuations with a wide range of characteristic B-scales in the vicinity of the QHE plateaus. This is consistent with earlier reports (as old as 20 years for the pioneering one [33]) which showed that in mesoscopic systems electronic transmission in the QHE regime is much more complex than in macroscopic 2D systems and gives rise to a broad spectrum of magnetoresistance oscillations with pseudo-AB super-periods or sub-periods [34-36]. These pseudo-AB oscillations rapidly disappear with increasing temperature and do not survive temperatures above 1K [14], in contrast with the orthodox AB oscillations, which can be observed up to temperatures in excess of 10 K [10]. Therefore, the physics behind the pseudo and orthodox AB oscillations must be different.

To explain the presence of sub-period oscillations, a recent theory [37] invokes Coulomb blockade of electrons tunnelling between the conducting edge states transmitted along the borders of the QR and those forming a quantum Hall electron island located at the centre of the device. In turn, the super-period oscillations could be explained [37] by Coulomb blockade in a quantum Hall electron island of much smaller extent, actually smaller than the QR arm width, centred therefore somewhere in the QR itself.

Our detailed SGM study published elsewhere [14] gives direct, i.e. visual, confirmation of this theory and demonstrates how SGM is able to decrypt the complexity of the magneto-resistance of the QR in the QHE regime. In particular, each sub-period AB-like set can be ascribed to a set of concentric fringes seen in the SGM map [14] that are reminiscent of those seen in the scanning-probe images of quantum dots in the Coulomb blockade regime [6,7]. Indeed, Coulomb blockade in a quantum dot, respectively Coulomb island, produces fringes in the SGM images that correspond to isopotential lines located at constant distances from the dot, respectively island. The fringes observed in Fig. 5a are likely due to such an effect. As far as the super-period fluctuations are concerned, SGM images gives strong confirmation of the above proposed interpretation [37] by locating the exact centre of the small quantum Hall Coulomb island within the QR arm [14].

## 7. Conclusion

The few examples presented in this paper confirm that SGM is very powerful in imaging the electronic transport in various low-dimensional semiconductor devices and to reveal how electrons behave down there. It often gives valuable complementary view on phenomena that are usually considered within a macroscopic experimental scheme. The ability of locating precisely compressible Coulomb islands in a quantum Hall interferometer is illustrative of this claim. Although some attention must be paid to avoid possible artifacts, the broad applicability range of SGM makes it a powerful tool for the electron diagnose of nanodevices in the coherent regime of transport, or even in the quantum Hall regime. Therefore, more spectacular achievements can be expected in the future.

## Acknowledgments

B. H. is postdoctoral researcher with the Belgian FRS-FNRS and F. M. is funded by FNRS and FCT (Portugal) postdoctoral grants. This work has been supported by FRFC grant no. 2.4.546.08.F, and FNRS grant no 1.5.044.07.F, by the Belgian Science Policy (Interuniversity Attraction Pole Program IAP-6/42) as well as by the PNANO 2007 program of the Agence Nationale de la Recherche, France ("MICATEC" project). VB acknowledges the award of a "Chaire d'excellence" by the Nanoscience Foundation in Grenoble.

**Figure captions :**

**Figure 1:** Principle of scanning-gate microscopy. A low-frequency (1 to 2 kHz) small probe current (of typically 20 nA) $I_{applied}$ is applied to the device, here a quantum ring patterned from a buried 2DEG. The measurement of the voltage drop $V_{measured}$ across the device gives access to its conductance $G$. The AFM tip is biased with a voltage $V_{tip}$ of the order of a few tenths of volts up to a few volts and scanned at a distance $d$ of typically 20 nm above the device (or alternatively, the sample is scanned under the fixed tip). This modifies the conductance and allows imaging the conductance changes $\Delta G(x,y)$ due to the tip perturbation as function of the lateral tip position *(x,y)*. Optionally, a perpendicular magnetic field $B$ can be applied. $\Delta G(x,y)$ forms what is called a SGM image throughout the paper.

**Figure 2:** Topographic (a) and SGM (b) images (T= 4.2 K, zero magnetic field) of a $Ga_{0.3}In_{0.7}As$ quantum ring acquired with an elongated tip ($V_{tip}$= -1.4 V) that has been damaged after thorough scanning. The light lines are guides to the eye pointing towards the device geometry. The ring-shaped device is the same as the one studied in [10]: the inner and outer ring diameters are 210 nm and 610 nm, respectively.

**Figure 3:** SGM images (a-c) of the same $Ga_{0.3}In_{0.7}As$ quantum ring as in Fig. 2. The ring geometry is shown schematically by full lines (image size: 2 µm x 2 µm; T= 4.2 K; zero magnetic field). The ring is connected to the 2D electron reservoir, which is buried at 25 nm below the free surface by two upper and lower narrow constrictions. The carrier concentration and mobility at 4.2K are $2.0 \; 10^{16}$ cm$^{-2}$ and 100.000 cm$^2$V$^{-1}$s$^{-1}$, respectively. The images are acquired for different $V_{tip}$ indicated on the figures and are all Fourier filtered [10] to compensate for a slowly varying strong background which masks part of the interference pattern. (d) and (e) are two sequences of profile plots as function of the tip bias. Each horizontal line corresponds to a vertical average of the conductance map in regions (α) and (β) shown in Fig. 3c, respectively. The translation of interference fringes to the left in (d) and to the right in (e) is a direct manifestation of the tip-induced electrostatic Aharonov-Bohm effect.

**Figure 4:** Quantum mechanical simulation of a SGM experiment on a QR in the presence of positively-charged impurities. The outer diameter, inner diameter and opening width are 530 nm, 280 nm, and 120 nm, respectively. The effective mass is 0.04 $m_0$, with $m_0$ the free electron mass. The Fermi energy is $E_F$= 107.4 meV. (a) and (b) are the simulated images of the LDOS and conductance changes (in units of $G_0$, the quantum of resistance), respectively, calculated for the random distribution of positively-charged impurities shown in (c). In the simulation, the tip potential has a Lorentzian shape with a 10 nm range and an amplitude of $E_F/50$. Reprinted from [13].

**Figure 5:** Very low-temperature (100 mK) high-magnetic field SGM experiments on a wide QR supplied with lateral gates. Here, the outer diameter is 1.2 µm and the opening width is 400 nm (insert in (a)). The materials parameters are the same as in Fig. 3. (a) exhibits the QR magnetoresistance with a 5.5 V voltage applied to the lateral gates. (b) and (c) show SGM resistance images at 8 T with a gate voltage of 6.45 nd 6.85 V, respectively. (d) is a SGM image acquired at a gate voltage of 7.2 V, where the QR is driven in the quantum Hall regime. All SGM images are recorded with no bias applied to the tip.

**FIGURE 1**

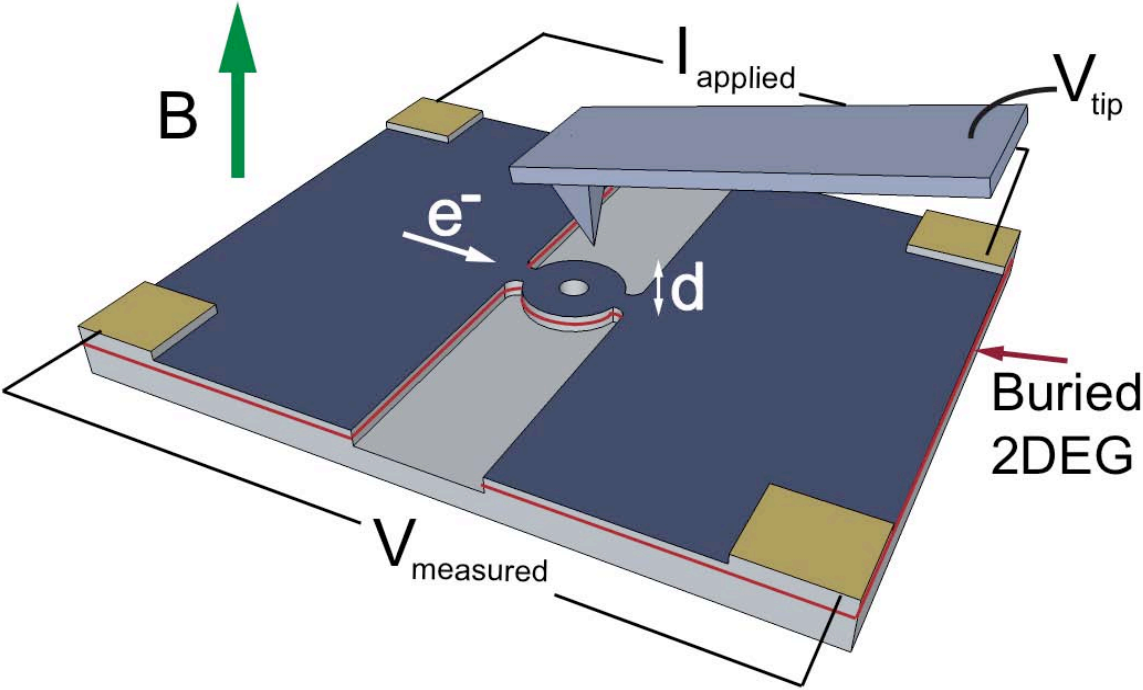

**FIGURE 2**

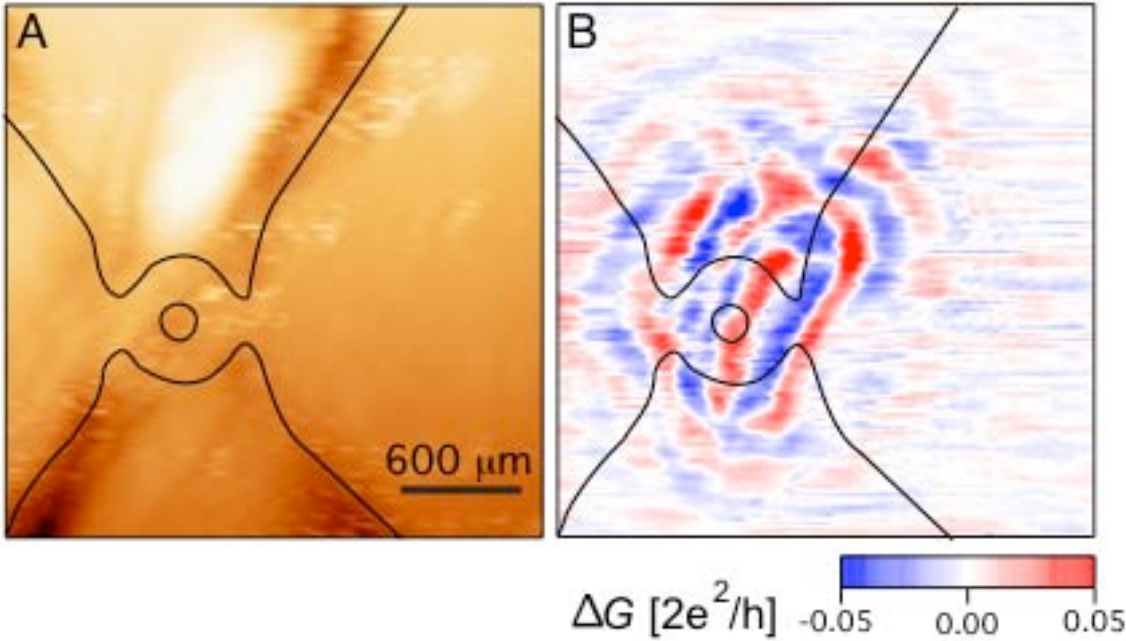

**FIGURE 3**

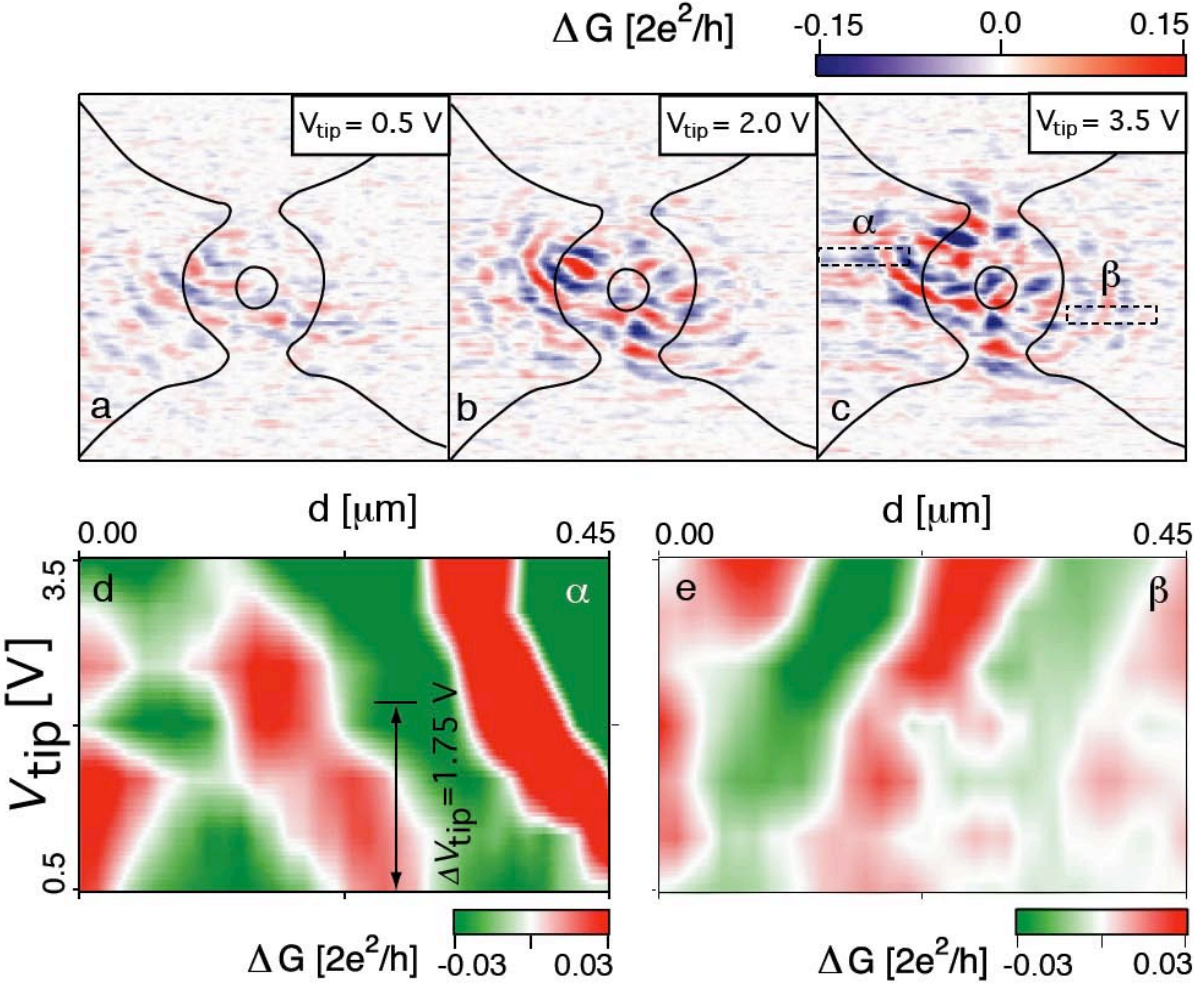

**FIGURE 4**

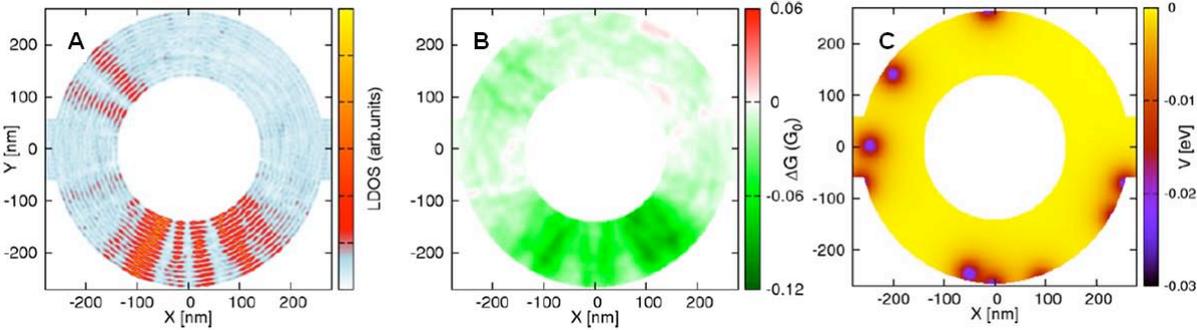

**FIGURE 5**

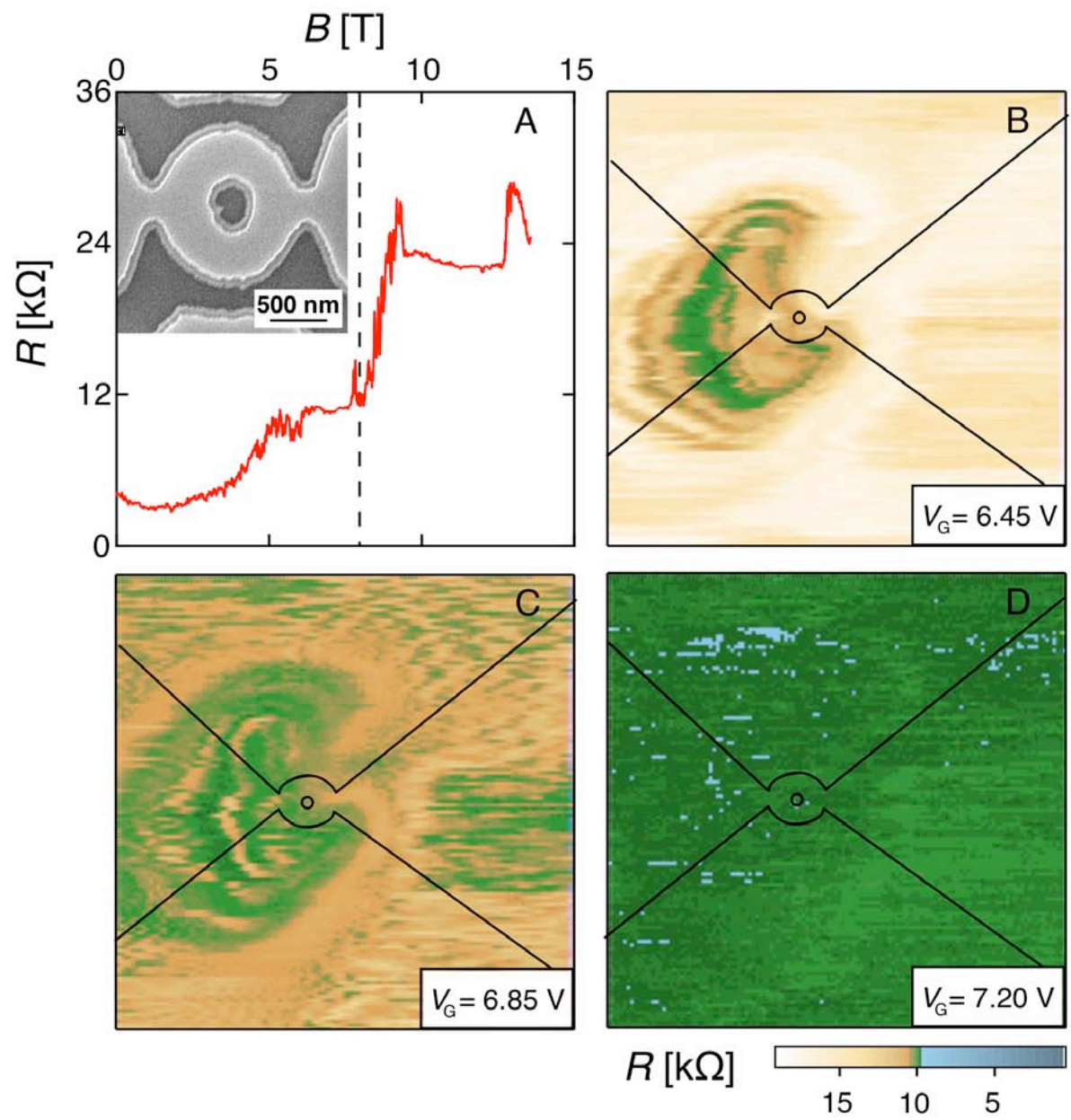